\documentclass[aps,showpacs,pra,twocolumn,amsmath]{revtex4}
\usepackage{epsfig}

\begin{document}

\title{Controllable spin-dependent transport in armchair graphene nanoribbon
structures}

\author{V. Hung Nguyen$^{1,2}$\footnote{E-mail: viet-hung.nguyen@u-psud.fr},
V. Nam Do$^3$, A. Bournel$^{1}$, V. Lien Nguyen$^2$ and P.
Dollfus$^{1}$}

\address{$^{1}$Institut d'Electronique
Fondamentale, UMR8622, CNRS, Universite Paris Sud, 91405 Orsay,
France \\ $^{2}$Theoretical Department, Institute of Physics,
VAST, P.O. Box 429 Bo Ho, Hanoi 10000, Vietnam \\
$^{3}$Hanoi Advanced School of Science and Technology, 1 Dai Co Viet
Str., Hanoi 10000, Vietnam}

\date{\today}
\begin{abstract}
Using the non-equilibrium Green's functions formalism in a tight
binding model, the spin-dependent transport in armchair graphene
nanoribbon (GNR) structures controlled by a ferromagnetic gate is
investigated. Beyond the oscillatory behavior of conductance and
spin polarization with respect to the barrier height, which can be
tuned by the gate voltage, we especially analyze the effect of
width-dependent band gap and the nature of contacts. The oscillation
of spin polarization in the GNRs with a large band gap is strong in
comparison with 2D-graphene structures. Very high spin polarization
(close to $100\%$) is observed in
normal-conductor/graphene/normal-conductor junctions. Moreover, we
find that the difference of electronic structure between normal
conductor and graphene generates confined states in the device which
have a strong influence on the transport quantities. It suggests
that the device should be carefully designed to obtain high
controllability of spin current.
\end{abstract}

\pacs{75.75.+a, 72.25.-b, 05.60.Gg, 73.43.Jn} \maketitle

\section{introduction}
Graphene, a monolayer of carbon atoms packed into a two dimensional
(2D) honeycomb lattice, has attached a great amount of attention
from both experimental and theoretical points of view \cite{cast09}
since it was isolated and demonstrated to be stable
\cite{novo04,novo05}. It is a basic building block for graphite
materials of all other dimensionalities, e.g., it can be wrapped up
into 0D fullerenes, rolled into 1D nanotubes, or stacked into 3D
graphite. Due to its unique electronic properties, i.e., its
conduction electrons behave as massless Dirac fermions
\cite{novo05,zhan05}, a lot of interesting phenomena such as the
finite conductance at zero concentration \cite{novo05}, the unusual
half integer quantum Hall effect \cite{zhan05}, and the Klein
tunneling \cite{stan09} have been observed in the graphene and
theoretically discussed in the framework of the massless fermion
Dirac's model \cite{novo05,kats06}.

Finite width graphene strips, which are referred as graphene
nanoribbons, have been also studied attentively
\cite{naka96,cres08,koba06,han007,quer08}. It has been shown that
GNRs of various widths can be obtained from graphene monolayers
using patterning techniques \cite{han007}. The transport properties
of the perfect GNRs are expected to depend strongly on whether they
have zigzag or armchair edges \cite{naka96}. In the framework of the
nearest neighbor tight binding (NNTB) model, the GNRs with zigzag
edges are always metallic while the armchair structures are either
semiconducting or metallic depending on their width. In the zigzag
GNRs, the bands are partially flat around the Fermi energy [$E = 0$
eV], which means that the group velocity of conduction electrons is
close to zero. Their transport properties are dominated by edge
states \cite{koba06}. In the GNRs with armchair edges, the bands
exhibit a finite energy gap in the semiconducting structures or are
gapless in the metallic ones \cite{naka96}. However, ab initio
studies have demonstrated that there are no truly metallic armchair
GNRs (see in Ref. \cite{cres08} and reference therein). Even for the
structures predicted to be metallic by the NNTB model, a small
energy gap opens, thus modifying their behavior from metallic to
semiconducting. In general, the group velocity of conduction
electrons in armchair GNRs is high, e. g., it is constant and equal
to about $10^6$ m/s \cite{novo05} in those metallic structures. The
transport properties of ideal zigzag and armchair GNRs are thus very
different. In the current work, we pay our attention only to the
ribbons with armchair edges.

Experimentally, electronic transport measurements through a graphene
sheet usually require contacts to metal electrodes, e.g., see an
illustration in Ref. \cite{huar07}. When tunneling from metal
reservoir to graphene occurs in a large area, the contact becomes
Ohmic and the area under the contact forms a substance which is a
hybrid between graphene and normal metal \cite{blan07}. Depending on
the nature of this substance, the system can be appropriately
considered as a graphene/graphene/graphene (GGG) structure or a
normal-conductor/graphene/normal-conductor (NGN) junction whose
contacts can be modeled by honeycomb or square lattices,
respectively. The ballistic transport through the NGN systems has
been investigated systematically in Refs.
\cite{scho07,blan07,robi07}.

Beside their interesting electronic transport properties, due to
very weak spin orbit interaction \cite{kane05}, which leads to a
long spin flip length ($\sim 1 \mu m$) \cite{jozs08}, the graphene -
based structures also offer a good potential for spin-polarized
electronics. Actually, graphene is not a natural ferromagnet.
However, recent works have shown that ferromagnetism and
spin-polarized states can be introduced in graphene, e. g., by
doping and by defect \cite{pere05,wehl08,yazy07} or by applying an
external electric field \cite{son006}. Especially, Haugen et. al.
\cite{haug08} suggested that the ferromagnetic correlations can be
created in graphene by the so-called proximity effect. The exchange
splitting induced by depositing a ferromagnetic insulator $EuO$ on
the graphene sheet was then roughly estimated to be about $5$ meV.
The effect has been also experimentally demonstrated
\cite{jozs08,hill06}. Motivated by these features, some other works
have predicted and discussed the controllability of spin current by
a ferromagnetic gate in $2D$-graphene structures
\cite{yoko08,vndo08,zou009}. The spin current was found to be an
oscillatory function of the potential barrier, which can be tuned by
the gate voltage, and its amplitude is never damped by the increase
of the width and the height of barrier \cite{yoko08}. However, the
spin polarization is not so high, e. g., its maximum value is just
about $30\%$. In addition, the other spin-dependent properties of
graphene such as spin field effect transistor \cite{seme07}, spin
Hall effects \cite{kane05,sini06}, spin valve effects
\cite{cho007,brey07,roja09,saff09} have been also investigated
extensively. Especially, the giant magneto-resistance has been
explored theoretically and discussed in the structures of armchair
\cite{saff09} and zigzag \cite{roja09} GNRs connecting two
conducting electrodes. The authors predicted that it can reach the
high value of $100\%$ \cite{roja09}.

In this article, we are interested in the possibilities of
electrically tunable of spin current in single ferromagnetic gate
armchair GNR structures. By investigating the physics of spin
polarized transport in these structures, we would like to derive
some simple scaling rules in order to tend to a high tunable spin
polarized current. The study is focussed on the role of the ribbon's
energy band gap and the different types of leads (either graphitic
or normal-conducting). In the NGN systems, since the strength of the
device-to-contact coupling and the device length are important
parameters, their influence on the control of spin current is also
carefully investigated. The paper is organized as follows. Section 2
is devoted to the description of the model and main formulas based
on the non-equilibrium Green's functions formalism (NEGF). In
section 3, the numerical results are presented and discussed.
Finally, a brief summary is given in section 4.

\section{model and formulation}
\begin{figure}[tbp]
\vspace{0.2cm}
\begin{center}
\epsfig{file=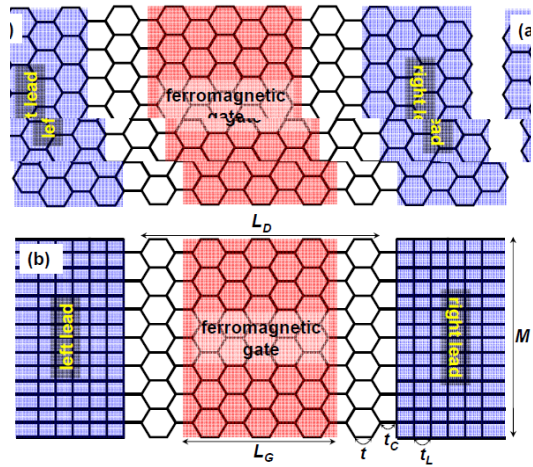,width=0.8\linewidth,angle=0,clip=}
\end{center}
\caption{(color online) Schematic illustration of the considered
armchair GNR structures with the number $M$ of carbon chains between
two edges: (a) graphitic and (b) normal-conducting leads. The latter
ones are modeled by square lattices. A magnetic gated insulator is
deposited to create a spin-dependent potential barrier in the center
of device.} \label{model}
\end{figure}

The considered structures consist of an armchair GNR coupled with
two electrodes which may be described either by graphitic (Fig.
1(a)) or normal-conducting (Fig. 1(b)) leads. In the simplest
consideration, the normal-conducting leads are modeled by square
lattices \cite{blan07,scho07,robi07}. A ferromagnetic gate is
assumed to create a potential barrier which controls the Fermi level
locally and to induce an exchange splitting into the device. To
model the structures, we use the single band tight binding
Hamiltonian
\begin{equation}\label{hamil}
\hat H = \hat H_L  + \hat H_D  + \hat H_R  + \hat H_C
\end{equation}
where $\hat{H}_{L,R}$ are the Hamiltonian of the left and right
leads, respectively; $\hat{H}_D$ is the Hamiltonian of the device;
$\hat{H}_C$ describes the coupling of the device to the leads. The
Hamiltonian terms in Eq. (1) can be written as
\begin{eqnarray}
\hat H_\alpha &=& \varepsilon _\alpha  \sum\limits_{i_\alpha ,\sigma
} {c_{i_\alpha  ,\sigma }^\dag  c_{i_\alpha  ,\sigma } }  - t_L
\sum\limits_{\left\langle {i_\alpha  ,j_\alpha  } \right\rangle
,\sigma } {c_{i_\alpha  ,\sigma }^\dag  c_{j_\alpha  ,\sigma } }
\nonumber \\ \hat H_D &=& \sum\limits_{i_d ,\sigma } { {\varepsilon
_{i_d, \sigma} } a_{i_d ,\sigma }^ \dag a_{i_d ,\sigma } }  -
t\sum\limits_{\left\langle {i_d ,j_d }
\right\rangle ,\sigma } {a_{i_d ,\sigma }^\dag  a_{j_d ,\sigma } }  \\
\hat H_C &=&  - t_C \sum\limits_{\alpha  = \left\{ {L,R} \right\}}
{\sum\limits_{\left\langle {i_\alpha ,j_d } \right\rangle ,\sigma }
{\left( {c_{i_\alpha  ,\sigma }^\dag  a_{i_d ,\sigma }  + h.c.}
\right)} } \nonumber
\end{eqnarray}
where the operators $c_{i_\alpha,\sigma}^\dag$
($c_{i_\alpha,\sigma}$) and $a_{i_d,\sigma}^\dag$ ($a_{i_d,\sigma}$)
create (annihilate) an electron with spin $\sigma$ in the electrode
$\alpha$ and the device region, respectively. The sum over carbon
atoms $\langle i,j \rangle$ is restricted to the nearest neighbor
atoms. $t$, $t_L$ and $t_C$ stand for the hopping parameters in the
device, the lead and at the coupling interface, respectively.
$\varepsilon_\alpha$ is the on-site energy of the leads which acts
as a shift in energy. The device spin-dependent on-site energy
$\varepsilon _{i_d, \sigma}$ is modulated by the gate voltage
\begin{equation}\label{barrier}
\varepsilon _{i_d ,\sigma }  = \left\{ \begin{array}{l}
 U_G  - \sigma h \,\,\,\,\,\,\,\,\,\,\,\,\, {\rm{in} \,\, \rm{gated} \,\,
 \rm{region}} \\
 0 \,\,\,\,\,\,\,\,\,\,\,\,\,\,\,\,\,\,\,\,\,\,\,\,\,\,\,\,\,\,\,\,\,
 {\rm{otherwise}}
 \end{array} \right.
\end{equation}
Here, $U_G$ denotes the potential barrier height, $h$ is the
exchange splitting and $\sigma = \pm 1$ describes the up/down spin
states.

Since no spin flip process is considered here, (1) can be decoupled
into two linear spin-dependent Hamiltonians $\hat H_\sigma$ and the
transport is easily considered using the NEGF formalism. For each
spin channel $\sigma$, the retarded Green's function is defined as
\begin{equation}\label{green}
\hat G_\sigma ^r \left( E \right) = \left[ {E + i0^+   - \hat
H_{D,\sigma }  - \hat \Sigma _L^r  - \hat \Sigma _R^r } \right]^{ -
1}
\end{equation}
where, $\hat \Sigma _\alpha^r$ describes the retarded self energy
matrices which contain the information on the electronic structure
of the leads and their coupling to the device. It can be expressed
as $\hat \Sigma _\alpha ^r  = \hat \tau _{{\rm{D}}{\rm{,}}\alpha }
\hat g_\alpha  \hat \tau _{\alpha ,D}$ where $\hat \tau$ is the
hopping matrix that couples the device to the leads. $\hat g_\alpha$
are the surface Green's functions of the uncoupled leads, i.e., the
left or right semi-infinite electrodes. The surface Green's
functions and the device Green's functions are calculated using the
fast iterative scheme \cite{sanc84} and the recursive algorithm
\cite{anan08}, respectively.

Within the model described by Eq. (1), the transport is considered
to be ballistic and the conductance through the device is calculated
using the Landauer formalism \cite{imry99}. The spin-dependent
conductances $\mathcal{G}_\sigma$ at the Fermi energy $E_F$ are
obtained from the transmission function $T_\sigma (E)$, such that
\begin{equation}\label{cond}
\mathcal{G}_\sigma \left( E_F \right) = \frac{{e^2 }}{h}T_\sigma
\left( E_F \right)
\end{equation}
and
\begin{equation}\label{trans}
T_\sigma  \left( E \right) = {\rm{Tr}}\left[ {\hat \Gamma _L \hat
G_\sigma ^r \hat \Gamma _R \hat G_\sigma ^a } \right]\cdot
\end{equation}
Here, $\hat G_\sigma ^a $ ($\equiv \hat G_\sigma ^{r\dag}$) denotes
the advanced Green's function. The tunneling rate matrix $\hat
\Gamma _{L\left( R \right)}$ for the left (right) lead is obtained
from
\begin{equation}\label{coupl}
\hat \Gamma _{L/R}  = {\rm{i}}\left[ {\hat \Sigma _{L/R}^r  - \hat
\Sigma _{L/R}^a } \right]
\end{equation}
where $\hat \Sigma _\alpha^a $ ($\equiv \hat \Sigma_\alpha
^{r\dag}$) is the advanced self energy. Finally, the spin
polarization is determined by
\begin{equation}\label{spin}
P = \frac{{\mathcal{G}_ \uparrow   - \mathcal{G}_ \downarrow
}}{{\mathcal{G}_ \uparrow + \mathcal{G}_ \downarrow }}\cdot
\end{equation}
In addition, the local density of states (LDOS) at site $j$ can be
also directly extracted from the retarded Green's function as:
\begin{equation}
{\rm{LDOS}}_{\left( j \right)} =  - \frac{1}{\pi }{\mathop{\rm
Im}\nolimits} G^r \left( {j,j} \right)\cdot
\end{equation}

By using the recursive algorithm described in Ref. \cite{anan08},
the size of matrices equals the number $M$ of carbon chains between
two edges. Thus, the cost of calculations is only linearly dependent
on the device length.

\section{results and discussion}

Using the above formalism, we investigate the spin-dependent
transport in the considered structures. Throughout the work, we set
$t = 2.66$ eV \cite{chic96} in the graphitic regions and assume that
$t_L$ is equal to $t$ and $t_C \leq t$ in the NGN junctions
\cite{blan07,robi07}. Our consideration is restricted to the
low-energy regime when $E \ll t$.

\subsection{Gate control of spin current in GGG structures}
\begin{figure}[tbp]
\vspace{0.2cm}
\epsfig{file=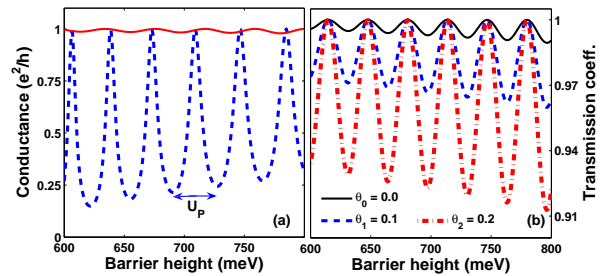,width=0.9\linewidth,angle=0,clip=}
\caption{(color online) (a) Oscillation of conductance vs the
barrier height $U_G$ in the GGG structures with different widths: $M
= 21$ (dashed) and $23$ (solid line). (b) illustrates the
transmission coefficient calculated from eq. (10) for different
modes $\theta_j$. Other
parameters are: $L_G = 42.5$ nm, $E_F = 300$ meV and $h = 0$ meV.}
\label{GNRC}
\end{figure}

As mentioned above, the gate voltage creates a potential barrier in
the device. In the armchair GNRs, the electronic properties of the
structures with $M \neq 3n + 2$ and $M = 3n + 2$ ($n$ is an integer)
are significantly different, i. e., there is a finite energy band
gap in the former structures while it is negligible in the latter
ones \cite{cres08}. Respectively, we display in Fig. 2(a) the
conductance as a function of the barrier height $U_G$ for the
structures: $M = 21$ (dashed) and $23$ (solid line). Here, the gated
region is assumed to be nonmagnetic, i. e., $h = 0$ meV. First of
all, the results show that the conductance has an oscillatory
behavior with respect to $U_G$. This phenomenon has been observed in
the $2D$-graphene structures and explained as a consequence of the
well-known Klein's tunneling \cite{kats06,yoko08}. In the framework
of the Dirac's description, the conductance peaks have been
demonstrated to be essentially due to the resonance or the good
matching of electron states and confined hole states outside/inside
the barrier region \cite{vndo08}, respectively. Those states, in
this model, correspond to electron states in positive and negative
energy bands. Note that due to the finite ribbon width, the
transverse momentum is quantized into a set of discrete values.
Practically, the oscillation of conductance can be seen clearly from
the analytical expression of the transmission coefficient for a
given transverse momentum mode $k_y^j$ (see the calculation in Ref.
\cite{klym08}). In the limit of low energy, it can be rewritten as
\begin{equation}\label{klein}
T = \frac{{\cos ^2 \theta_j \sin ^2 {\phi} }}{{\cos ^2 \theta_j \sin
^2 {\phi} + \left( {\sin \theta_j + \cos {\phi} } \right)^2 \sin ^2
\left( {k_x^b L_G} \right)}}
\end{equation}
where $\theta_j = \tan^{-1}(k_y^j/k_x)$ and $\sin \phi  = \left[
{\left( {t - {\rm{v}}_F k_y^j} \right)\sin \left( {3ak_x^b /2}
\right)} \right]/\left( {U_G  - E} \right)$ with ${\rm v}_F = 3at/2$
and $a$ is the $C-C$ bond length. $k_x$ ($k_y^j$) denotes the
longitudinal (transverse) momentum, which is the deviation of the
momentum $\vec k$ from the zero energy point, outside the barrier
and $k_x^b$ is the longitudinal momentum inside the barrier. The
energy dispersions outside/inside the barrier are, respectively
\begin{eqnarray}
 E &=& {\rm{v}}_{\rm{F}} \sqrt {k_x^2  + k_y^{j2} }  \\
 E - U_G &=& - \sqrt {4t\left( {t - {\rm{v}}_F k_y^j } \right)\sin ^2
\frac {3ak_x^b}{4} + {\rm{v}}_F^2 k_y^{j2} }
\end{eqnarray}
Accordingly, the transmission and then the conductance have their
maximum (or minimum) values when $k_x^bL_G$ is equal to $m\pi$ (or
$(m+1/2)\pi$) for any integer $m$. In the limit of $E \ll U_G \ll
t$, the eq. (12) can be rewritten as $U_G - E \approx {\rm v}_F
k_x^b$ and the period of oscillation is defined by $U_P = {\rm
v}_F\pi/L_G$, which coincides with that in Refs.
\cite{yoko08,vndo08}. In general cases, when the relation between
$U_G$ and ${\vec k}$ is nonlinear, the period can be approximately
expressed as $U_P = {\rm v}_g\pi/L_G$ with ${\rm v}_g \leq {\rm
v}_F$. For instance, ${\rm v}_g \approx 0.74 {\rm v}_F$ and
$0.89{\rm v}_F$ for $M  = 21$ and $23$ presented in Fig. 2(a),
respectively.

As a consequence of different electronic structures, the Fig. 2(a)
also shows that the conductance for the case of $M = 21$ (large
energy gap) oscillates strongly in comparison with the other ($M =
23$). This can be easily understood by considering the behavior of
transmission coefficient for different energy gaps. As seen in Eq.
(11), the energy gap is enlarged when increasing $k_y^j$ (or
$\theta_j$). So that, from the Eq. (10) and the Fig. 2(b), we see
that when the energy gap is larger (increasing $\theta_j$), the
oscillation of transmission is stronger. It leads to the different
behaviors of conductance shown in Fig. 2(a). Similarly, it is shown
that the oscillation of conductance in the structure with $M = 21$
is also stronger than that in the $2D$-graphene structures (see in
Fig. 2(a) of Ref. \cite{yoko08}), where the gap is truly zero.

In eqs. (10-11), the mode $\theta_0 = 0$ (or $k_y^0 = 0$)
corresponds essentially to the normal incident mode whose energy
dispersion is gapless and linear in the $2D$-graphene structures.
When considering the behavior of transmission coefficient, we find
an important feature: it is not uniformly equal to unity, but a
function of the barrier height even in the case of $\theta =
\theta_0$. This differs from the prediction of Klein's paradox
obtained by using Dirac's description in graphene structures
\cite{kats06} as previously discussed in Ref. \cite{tang08}.
Practically, the transmission coefficient (eq. (10)) approaches the
simplified expression (4) in Ref. \cite{kats06} only in the limit of
$E \ll U_G \ll t$.
\begin{figure}[tbp]
\vspace{0.2cm}
\epsfig{file=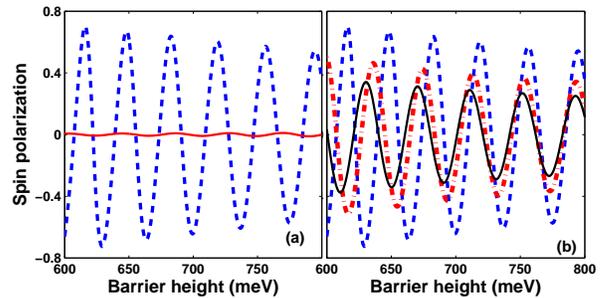,width=0.9\linewidth,angle=0,clip=}
\caption{(color online) (a) Spin polarization as a function of the
barrier height $U_G$ for the same structures as in Fig. 2(a). (b)
shows an example of the effects of the different ribbon widths on
the spin polarization: $M = 21$ (dashed), $27$ (dashed-dotted) and
$33$ (solid line). Everywhere $L_G = 42.5$ nm, $E_F = 300$ meV and
$h = 10$ meV.} \label{GNRC}
\end{figure}

Now we investigate the behavior of spin polarization in the
ferromagnetic gate structures. The exchange splitting $h$ is chosen
to be $10$ meV, which can be achieved experimentally
\cite{jozs08,hill06}. Since no spin flip process is considered, the
exchange splitting just shifts the conductance of each spin channel
relatively to the other. The spin polarization therefore behaves as
an oscillatory function of $U_G$ as shown in Fig. 3. Similar
phenomena in the $2D$-graphene structures have been also observed
and discussed in Ref. \cite{yoko08,vndo08,zou009}. It was shown that
the oscillation of spin current is never damped with the increase of
the width and the height of barrier and the spin polarization can be
reversed by changing the gate voltage. Actually, the amplitude of
$P$ depends primarily on the phase coherence/decoherence of the
oscillation of spin - dependent conductances, i. e., it has the
maximum/minimum value when the gate length (or the barrier width)
$L_G$ is equal to a half-integer/integer of $L_h$ with $L_h = {\rm
v}_g\pi/2h$, respectively. Hence, the gate control of spin current
can be modulated by changing $L_G$, i. e., it leads to the beating
behavior of $P$ similar to that shown in Fig. 5(c) of Ref.
\cite{vndo08}. Furthermore, as a consequence of the behavior of
conductance presented in Fig. 2(a), Fig. 3(a) also shows that the
oscillation of $P$ in the GNRs with a large energy gap is very
strong in comparison with the others. For instance, the amplitude of
$P$ is about $65\%$ for $M = 21$ while it is only few percents for
$M = 23$ or has a maximum value of $30\%$ in the $2D$-graphene
structures \cite{yoko08,vndo08,zou009}. However, since for $M \neq
3n + 2$ the energy gap decreases when increasing the ribbon width,
the oscillation of conductance and $P$ in those structures is
gradually weaker. The transport quantities for the GNRs in the limit
of infinite width approaches those for the $2D$-graphene structures,
where the continuum Dirac's description is valid. To illustrate this
point, we display in Fig. 3(b) an example of the effect of different
ribbon widths on the spin polarization in the structures with $M
\neq 3n + 2$. Indeed,  when increasing the ribbon width, the
amplitude of $P$ decreases and becomes closer to that in
2D-graphene, i. e., it is only about $35\%$ for $M = 33$.

\subsection{Effects of normal-conducting leads}

In this section, we consider the spin transport in NGN junctions.
First, we focus on the possibilities of obtaining high tunable spin
current when replacing the graphitic leads by the normal-conducting
ones. Second, we analyze the sensitivity of transport quantities to
different parameters as the device length and the Fermi energy.

In Ref. \cite{scho07}, Schomerus compared the resistances of NGN
junctions and GGG structures, and found the duality between
graphitic and normal-conduncting contacts. He has shown that
identical transport properties arise when the graphitic leads are
replaced by quantum wires and the difference between the results
obtained in those structures is only quantitative. On this basis, we
plot in Fig. 4(a) the conductance as a function of $U_G$ in the NGN
junction in comparison with the GGG structure for the case of $L_D =
51$ nm and $E_F = 300$ meV. Since we found that the results depend
weakly on the hopping energy in the leads, $t_L$ is chosen to be
equal to $t$ for simplicity. Qualitatively, Fig. 4(a) shows that the
oscillation of conductance seem to be unchanged in its phase and
period when changing the leads. Quantitatively, the oscillation in
the NGN junction is stronger than in the GGG structure. This can be
explained clearly by the effect of replacing the graphitic leads by
the normal conducting ones on the picture of bound states in the
barrier region. These states, in the framework of the Dirac's
description, have been considered as confined hole states in the
$2D$-graphene structures \cite{vndo08}. In Fig. 4(b), we display the
LDOS (see the right axis for graphitic and the left axis for
normal-conducting leads) at the first site of barrier region with
respect to $U_G$. The results are understood that the peaks of LDOS
occur when the Fermi energy corresponds to any bound state.
Obviously, the oscillation of LDOS (or the quantization of bound
states in the barrier) in the NGN junction appears stronger (with
higher peaks) than that in the GGG structure. It is the essential
origin of the different behaviors of conductance as shown in Fig.
4(a).
\begin{figure}[tbp]
\vspace{0.2cm}
\epsfig{file=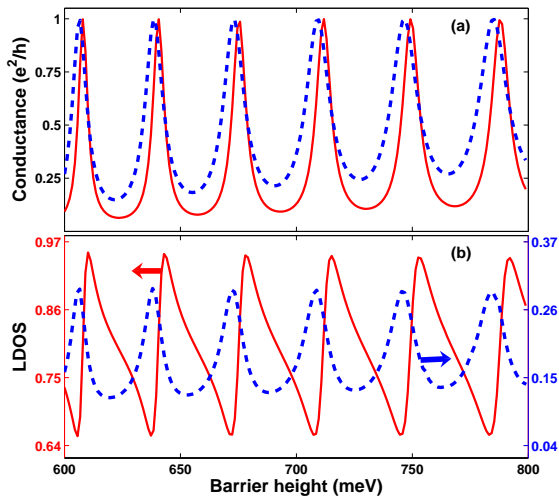,width=0.85\linewidth,angle=0,clip=}
\caption{(color online) Comparison of conductance (a) and LDOS (b)
in different structures: graphitic (dashed) and normal-conducting
(solid lines, $t_C = t$) leads. Everywhere $M = 21$, $L_D = 51$ nm,
$L_G = 42.5$ nm, $E_F = 300$ meV and $h = 0$ meV.} \label{CONTE}
\end{figure}

Now we turn to the behavior of spin current in ferromagnetic gate
NGN junctions, i. e., $h = 10$ meV. In Figs. 5(a) and 5(b), we
display the comparison of spin polarization in the NGN junctions and
the GGG structures. Due to the different behaviors of conductance
shown in Fig. 4(a), the amplitude of $P$ in former structures is
remarkably larger than that in latter ones. Particularly, when
changing the leads, it increases from $65\%$ to $81\%$ (see in Fig.
5(a)) and from $1\%$ to $50\%$ (see in Fig. 5(b)) for $M = 21$ and
$23$, respectively. Moreover, in the NGN junctions, the possibility
of obtaining high tunable spin current is more impressive with
decreasing the strength of the device-to-contact coupling, which is
characterized by the hopping energy $t_C$. In Figs. 5(c) and 5(d),
we plot the obtained results for three cases: $t_C = t$ (dotted),
$0.8t$ (dashed) and $0.6t$ (solid lines). Actually, the transport in
these structures depends strongly on the properties of junctions and
therefore on $t_C$. A smaller $t_C$ corresponds to a higher contact
resistance \cite{blan07}. We find that the quantization of bound
states in the barrier region is stronger when decreasing $t_C$ (not
shown), which leads to a stronger oscillation of the transport
quantities with respect to the barrier height. Indeed, even in the
case of $M = 23$, the spin polarization can reach a very high value
of $86\%$ for $t_C = 0.6t$ (see in Fig. 5(d)). More impressively, in
the cases of $M = 21$, it can tend to $100\%$ by reducing $t_C$ (see
in Fig. 5(c)). A similar feature (giant magneto resistance) has been
also predicted in the structures of GNRs connecting two conducting
electrodes \cite{roja09,saff09}.
\begin{figure}[tbp] \vspace{0.2cm}
\epsfig{file=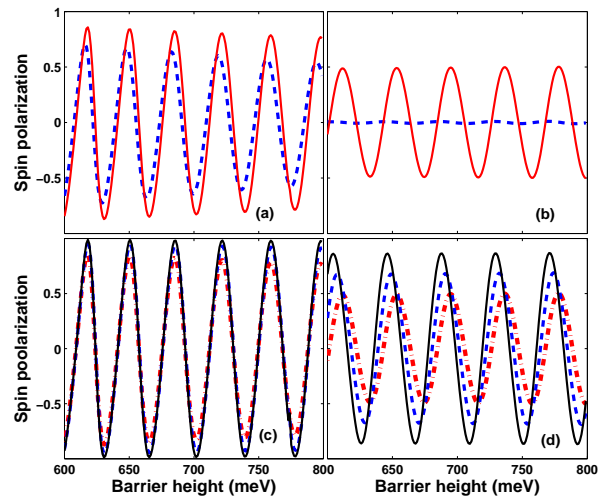,width=0.9\linewidth,angle=0,clip=}
\caption{(color online) (a,b) Comparison of spin polarization in the
different structures: graphitic (dashed) and normal-conducting
(solid lines, $t_C = t$) leads. (c,d) The spin polarization in the
latter one with different coupling strengths: $t_C = t$ (dashed),
$0.8t$ (dashed-dotted) and $0.6t$ (solid lines). The ribbon widths
are $M = 21$ (a,c) and 23 (b,d). Other parameters are $L_D = 51$ nm,
$L_G = 42.5$ nm, $E_F = 300$ meV and $h = 10$ meV.} \label{CONTE}
\end{figure}

Practically, the features discussed above depend strongly on the
parameters of the NGN junctions, such as the device length and/or
the Fermi energy. It results from the fact that the charge transport
can be confined in the device by two normal-conductor/graphene
junctions. It leads to an additional resonant condition controlling
the transport picture beside the transmission via the bound states
in the barrier. Indeed, the existence of such confined states is
demonstrated clearly from the behavior of LDOS shown in Fig. 6(a).
In this figure, to cancel the effects of bound states in the
barrier, the gate voltage is not applied to the device. The energy
spacing $E_S$ is estimated to be about $25.5$ meV for $L_D = 51$ nm
and $12.6$ meV for $102$ nm. It means that $E_S$ seems to be
inversely proportional to the device length, i. e., as illustrated
in Fig. 6(c). This implies an unusual quantization of charges in the
graphene-based structures, which is essentially different from the
case of normal semicondutors wherein $E_S \propto 1/L_D^2$ as
previously discussed in Refs. \cite{vndo08,milt06}. Due to such
confinement, the transport quantities, such as the conductance and
the spin current (not shown), have an oscillatory behavior also with
respect to the Fermi energy in the considered region (see in Fig.
6(b)). Therefore, when a gate voltage is applied, there is a
coexistence of bound states in the barrier and confined states in
the device. They together respond for the resonant transport
conditions of the structure.
\begin{figure}[tbp]
\vspace{0.2cm}
\epsfig{file=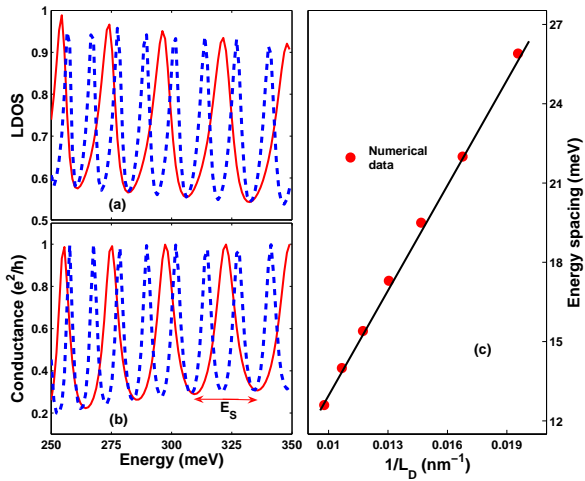,width=0.89\linewidth,angle=0,clip=}
\caption{(color online) (a) LDOS illustrating the existence of
confined states in the device and (b) conductance in the NGN
junctions as a function of the Fermi energy for different device
lengths: $L_D = 51$ nm (solid) and 102 nm (dashed lines). (c) shows
the dependence of energy spacing of the confined states on the
inverse of device length. Everywhere $M = 21$, $L_G = 42.5$ nm, $t_C
= 0.8t$ and $U_G = h = 0$ meV.} \label{SIZEE}
\end{figure}
\begin{figure}[tbp]
\vspace{0.2cm}
\epsfig{file=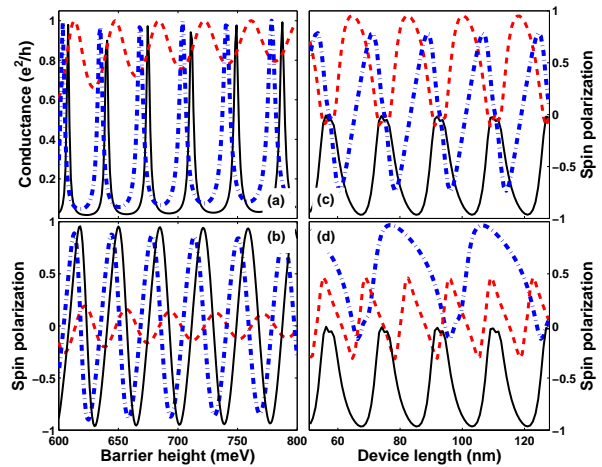,width=0.9\linewidth,angle=0,clip=}
\caption{(color online) (a) Conductance and (b) spin polarization
$P$ in the NGN structures as functions of the barrier height $U_G$
for different device lengths: $68$ nm (solid), $74$ nm (dashed) and
$79$ nm (dashed-dotted lines). The oscillation of $P$ versus the
device length for different values of $U_G$ (c): $630$ meV (solid),
$655$ meV (dashed-dotted) and $681$ meV (dashed line); and $E_F$
(d): $250$ meV (dashed-dotted), $300$ meV (solid) and $350$ meV
(dashed line). Other parameters are $M = 21$, $L_G = 42.5$ nm, $E_F
= 300$ meV (in (a,b,c)), $t_C = 0.8t$, $U_G = 630$ meV (in (d)) and
$h = 10$ meV.} \label{SIZEE}
\end{figure}

On the other hand, the gate controllability of spin current is
principally due to the picture of bound states in the barrier (or
Klein's tunneling) \cite{vndo08}. It arises a question about how the
confined states in the device affect that picture. As shown in Figs.
4 and 5, the replacement of the graphitic leads (infinite $L_D$) by
normal-conducting ones (finite $L_D$) does not affect the period,
but the amplitude of the oscillation. It suggests that, in the NGN
junctions, the oscillation of conductance and spin polarization can
be modulated in its amplitude while its period is unchanged when
changing the device length. To examine this statement, we plot the
conductance in Fig. 7(a) and the spin polarization in Fig. 7(b) as
functions of the barrier height for different device lengths. From
Fig. 7(a), we see that while the period is determined only by the
gate length, the oscillation of conductance is modified by changing
$L_D$, i. e., it is strong/weak when $L_D = 68$ (and $79$) or $74$
nm, respectively. Consequently, the amplitude of spin polarization
is dependent on $L_D$, i. e., it is about $95\%$ for $L_D = 68$ nm,
$15\%$ for $74$ nm and $86\%$ for $79$ nm (see Fig. 7(b)). This
feature is exhibited more clearly in Fig. 7(c) by three curves of
spin polarization versus $L_D$ for different barrier heights: $U_G =
630$ (dashed), $655$ (dashed-dotted), and 681.5 (solid line) meV. We
see that the spin current has an oscillatory behavior and is
suppressed completely at certain values of $L_D$. Obviously, this
demonstrates that the amplitude of spin polarization exhibited in
Fig. 7(b) is also an oscillatory function of the device length.
Moreover, its period seems to be inversely proportional to the Fermi
energy, i. e., it is about $27.8$ nm for $E_F = 250$ meV, $18.3$ nm
for $300$ meV and $13.4$ nm for $350$ meV (see Fig. 7(d)). It is
nothing, but a consequence of the resonant transport due to the
confined states in the device. Hence, the gate control of spin
current in the NGN junction can be modulated not only by the gate
length $L_G$ (see in the section A) but also by the device length
$L_D$ and/or the Fermi energy $E_F$. This implies that the structure
should be carefully designed to obtain high controllability of spin
current.

\section{conclusions}

Using the NEGF method for quantum transport simulation within a
tight binding hamiltonian, we have considered the spin-dependent
transport in single ferromagnetic gate armchair GNR structures. The
leads are modeled as either graphitic or normal-conducting.

In the case of graphitic leads, it is shown that the conductance and
the spin current behave as oscillatory functions of barrier height
which can be tuned by the gate voltage. The oscillation of spin
polarization in the ribbon structures with a large energy band gap
is strong in comparison with the $2D$-graphene structures.
Especially, the study has demonstrated that a very high spin
polarization can be observed in the NGN junctions. It results from
the fact that the quantization of bound states in the barrier
(gated) region can appear very strong when using the
normal-conducting leads. In this structure, it is shown that the
spin polarization increases and can tend to $100\%$ with the
decrease of the strength of the device-to-contact coupling.
Moreover, we have also found the existence of confined states in the
device by normal-conductor/graphene junctions. This confinement
responds for an additional resonant condition beside the
transmission via the bound states in the barrier. Therefore, the
gate control of spin current in the NGN junctions can be modulated
not only by the gate length but also by the device length and/or the
Fermi energy.

Our predictions may be helpful for designing efficient spintronics
devices based on perfect armchair GNRs. However, some disorder
effects, e. g. due to edge roughness, have been observed
experimentally \cite{han007} and demonstrated to affect the
transport properties \cite{quer08} of the GNR structures. Further
work is needed to assess their influence on the spin polarized
properties discussed in this article.

\textbf{Acknowledgements.} This work was partially supported by the
European Community through the Network of Excellence NANOSIL.

\end{document}